\documentclass[a4paper,11pt]{article}
\pdfoutput=1

\usepackage{jinstpub}

\usepackage{lineno}
\usepackage{graphicx}
\usepackage{dcolumn}
\usepackage{bm}
\usepackage{hyperref}
\hypersetup{colorlinks=true, citecolor=blue, urlcolor=blue, linkcolor=blue}
\usepackage{siunitx}
\usepackage{xcolor}
\usepackage[caption=false]{subfig}

\DeclareSIUnit\angstrom{\text {Å}}

\title{Channeling Acceleration in Crystals and Nanostructures and Studies of Solid Plasmas: New Opportunities}

\author[1]{Max F. Gilljohann,}
\author[1]{Yuliia Mankovska,} 
\author[1]{Pablo San Miguel Claveria,}
\author[2,3]{Alexei Sytov,}
\author[2]{Laura Bandiera,}
\author[4,5]{Robert Ariniello,} 
\author[6,7]{Xavier Davoine,}
\author[5]{Henrik Ekerfelt,}
\author[5]{Frederico Fiuza,} 
\author[6,7]{Laurent Gremillet,}
\author[1]{Alexander Knetsch,}
\author[8]{Bertrand Martinez,}
\author[1]{Aimé Matheron,}
\author[9]{Henryk Piekarz,} 
\author[5]{Doug Storey,}
\author[10]{Peter Taborek,} 
\author[10]{Toshiki Tajima,}
\author[9]{Vladimir Shiltsev,}
\author[1]{Sébastien Corde}

\affiliation[1]{LOA, ENSTA Paris, CNRS, Ecole Polytechnique, Institut Polytechnique de Paris, 91762 Palaiseau, France}
\affiliation[2]{INFN Ferrara Division, Via Saragat 1, 44122 Ferrara, Italy}
\affiliation[3]{Korea Institute of Science and Technology Information (KISTI), 245 Daehak-ro, Yuseong-gu, Daejeon 34141, Korea}
\affiliation[4]{University of Colorado Boulder, Department of Physics, Center for Integrated Plasma Studies, Boulder, Colorado 80309, USA}
\affiliation[5]{SLAC National Accelerator Laboratory, Menlo Park, CA 94025, USA}
\affiliation[6]{CEA, DAM, DIF, 91297 Arpajon, France}
\affiliation[7]{Universit\'{e} Paris-Saclay, CEA, LMCE, 91680 Bruy\`{e}res-le-Ch\^{a}tel, France}
\affiliation[8]{GoLP/Instituto de Plasmas e Fusa\~o Nuclear, Instituto Superior T\'ecnico, Universidade de Lisboa, 1049-001 Lisboa, Portugal}
\affiliation[9]{Fermi National Accelerator Laboratory, Batavia, IL 60510, USA}
\affiliation[10]{University of California Irvine, Irvine, CA 92697, USA}

\emailAdd{max.gilljohann@polytechnique.edu} \emailAdd{sebastien.corde@polytechnique.edu}

\abstract{
Plasma wakefield acceleration (PWFA) has shown illustrious progress and resulted in an impressive demonstration of tens of \si{\giga\electronvolt} particle acceleration in meter-long single structures. To reach even higher energies in the \SIrange{1}{10}{\tera\electronvolt} range, a promising scheme is channeling acceleration in solid-density plasmas within crystals or nanostructures. 

The \textit{E336} experiment studies the beam-nanotarget interaction with the highly compressed electron bunches available at the FACET-II accelerator. These studies furthermore involve an in-depth research on dynamics of beam-plasma instabilities in ultra-dense plasma, its development and suppression in structured media like carbon nanotubes and crystals, and its potential use to transversely modulate the electron bunch. 
}

\begin{document}
\maketitle
\flushbottom

\section{Introduction: plasma wakefields in solids (amorphous and structured)}

The feasibility of future colliders critically depends on their energy reach, luminosity, cost, length and power efficiency \cite{VSFZ2021}. So far, the most advanced proposals for energy frontier colliders call for acceleration by plasma wakefields. These can be excited by: lasers (demonstrated electron energy gain of about \SI{8}{\giga\electronvolt} over \SI{20}{\centi\meter} of plasma with \SI{3e17}{\per\cubic\centi\meter} density at the BELLA facility at LBNL~\cite{Gonsalves2019}); very short electron bunches (\SI{9}{\giga\electronvolt} gain over \SI{1.3}{\meter} of \SI{\sim e17}{\per\cubic\centi\meter} density plasma at the FACET facility at SLAC~\cite{Litos2016}) and by proton bunches (some \SI{2}{\giga\electronvolt} gain over \SI{10}{\meter} of \SI{e15}{\per\cubic\centi\meter} density plasma at the AWAKE experiment at CERN~\cite{Adli2018}). While  multi-\si{\tera\electronvolt} $e^+e^-$ colliders are thought to be accessible to plasma wakefield accelerators (PWFA), there is, however, a number of critical issues to be resolved along that path, such as the power efficiency of the laser/beam PWFA schemes; acceleration of positrons (which are focused differently when accelerated in plasma); efficiency of staging (beam transfer and matching from one short plasma accelerator cell to another); beam emittance control in scattering media; and beamstrahlung (radiation due to beam-beam interaction) that leads to an rms energy spread at IP of about \SI{30}{\percent} for a \SI{10}{\tera\electronvolt} high-luminosity collider and \SI{80}{\percent} for a \SI{30}{\tera\electronvolt} high-luminosity collider~\cite{Alegro2019,Barklow2023}.

Assessments of options for "ultimate" future energy frontier colliders with c.\,o.\,m.\ energies far beyond LHC's \SI{14}{\tera\electronvolt} (the most powerful modern accelerator) shows~\cite{VSFZ2021, VS2021} that for the same reason the circular $e^+e^-$ collider energies do not extend beyond the Higgs factory range (\SI{0.25}{\tera\electronvolt}), there will be no circular proton-proton colliders beyond \SI{100}{\tera\electronvolt} because of unacceptable synchrotron radiation power – therefore, the colliders will have to be linear. Moreover, electrons and positrons even in linear accelerators become impractical above about \SI{3}{\tera\electronvolt} due to growing facility site power demands, beamstrahlung at the IPs, and due to the radiation in the focusing channel above about \SI{10}{\tera\electronvolt}. If one goes further and requests such a flagship machine not to exceed \SI{\sim 10}{\kilo\meter} in length then an accelerator technology is needed to provide an average accelerating gradient of over \SI{30}{\giga\electronvolt\per\meter} (to be compared with an equivalent \SI{\sim 0.5}{\giga\electronvolt\per\meter} in the LHC). There is only one such option known now: super-dense plasma as, e.\,g., in crystals~\cite{TC1987}, which excludes protons because of nuclear interactions and leaves us with muons as the particles of choice. Indeed, the (conduction electron) density of charge carriers $n_0$ in solids ($\SI{e22}{\per\cubic\centi\meter} \lesssim n_0 \lesssim \SI{e22}{\per\cubic\centi\meter}$) is significantly higher than that in gaseous plasma, and correspondingly, the longitudinal accelerating fields of up to \SI{100}{\tera\electronvolt\per\meter} are possible according to 
\begin{equation}
E\left[\si{\giga\volt\per\meter}\right] = m_e \omega_p c/e  \approx 100 \sqrt {n_0 \left[\SI{e18}{\per\cubic\centi\meter}\right]} .
\label{grad}
\end{equation}
and allow envisioning compact \SI{1}{\peta\electronvolt} linear crystal muon colliders \cite{VS2012}. Another opportunity is the creation of muons within the plasma acceleration stage to generate ultra-low emittances, as proposed in Ref.~\cite{PWAsource2022}. 
The choice of muons is beneficial because of small scattering on solid media electrons, absence of beamstrahlung effects at the IP, and continuous focusing while channeling in crystals, i.\,e., the acceleration to final energy can be done in a single stage. Muon decay becomes practically irrelevant in such very fast acceleration gradients as muon lifetime quickly grows with energy as $\gamma \times \SI{2.2}{\micro\second}$. 

High luminosity cannot be expected for such a facility if the beam power $P$ is limited at, e.\,g., $P\leq \SI{100}{\mega\watt}$. In that case, the beam current will have to go down with the particle energy as $I=P/E$, and, consequently, the luminosity will by necessity go down with energy $E$. This trend can only be partially reset by formation of ultra-small \si{\angstrom} beam sizes at the collision points. Initial luminosity analysis of such machines assumes a small number of muons per bunch $\mathcal{O}(1000)$, a small number of bunches $\mathcal{O}(100)$, high repetition rate $\mathcal{O}(\SI{1}{\mega\hertz})$ and ultimately small sizes and overlap of the colliding beams $\mathcal{O}(\SI{1}{\angstrom})$ -- see Sec.~\ref{collider} below.

\begin{figure}[ht]
\begin{center}
\includegraphics[width=3.2in]{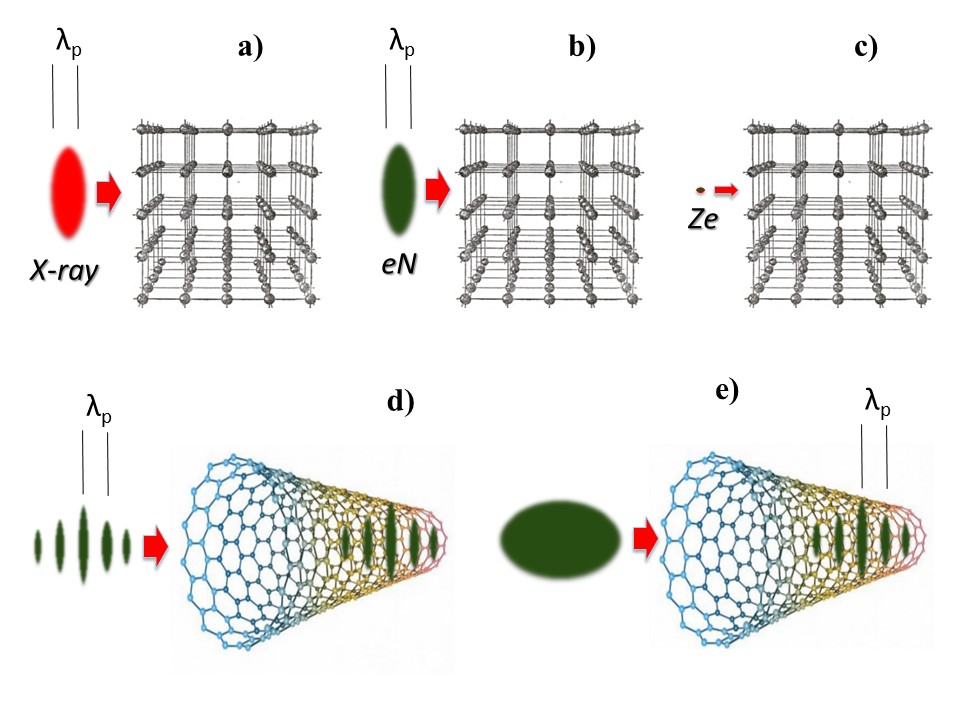}
\end{center}
\caption{Possible ways to excite plasma wakefields in crystals or/and nanostructures: a) by short X-ray laser pulses; b) by short high density bunches of charged particles; c) by heavy high-$Z$ ions; d) by modulated high current beams; e) by longer bunches experiencing self modulation instability in the media (from Ref. \cite{VS2019}).}
\label{drivers}
\end{figure}

In general, excitation of wakefields in crystals or nanostructures can be achieved with either short sub-\si{\micro\meter} high-density bunches of X-ray laser pulses or charged particles (electrons, high-$Z$ ions, etc.), or by pre-modulated or self-modulated very high current bunches -- see Fig.~\ref{drivers}. For example, bunches of charged particles can excite plasma effectively if their transverse and longitudinal sizes are comparable or shorter than the plasma wavelength $\lambda_p \sim \SI{0.3}{\micro\meter}$ for $n_0=\SI{e22}{\per\cubic\centi\meter}$ and the total number of particles approaches the number of free electrons in the solid plasma $\sim n_0 \lambda_p^3$.

There are several critical phenomena in the solid plasma due to intense energy radiation in high fields and increased scattering rates which result in fast pitch-angle diffusion over distances of $l[\si{\meter}] \simeq E[\si{\tera\electronvolt}]$. The scattering leads to particles escaping from the driving field; thus, it was suggested to accelerate particles (muons) in solids along the major crystallographic directions, which provides a channeling effect in combination with low emittances determined by an Angstrom-scale aperture of the atomic tubes \cite{CN1997,DF2008}. Channeling in the nanotubes was later brought up as a promising option \cite{TTPA1990, ZHANG2016, SHIN2013, SHIN2015}. Positively charged particles are channeled more robustly, as they get repelled from ions and, thus, experience weaker scattering. Radiation emission due to the betatron oscillations between the atomic planes is thought to be the major source of energy dissipation, and the maximum beam energies are limited to about \SI{0.3}{\tera\electronvolt} for positrons, \SI{10}{\peta\electronvolt} for antimuons and \SI{1000}{\peta\electronvolt} for protons \cite{CN1997}. For energies of \SIrange{1}{10}{\peta\electronvolt}, muons offer much more attraction because they are point-like elementary particles and, contrary to protons, do not carry an intrinsic energy spread of elementary constituents; and they can much easier propagate in solid plasma than protons which will extinct due to nuclear interactions. Very high gradient crystal/CNT accelerators might need to be disposable if the externally excited fields exceed the ionization thresholds and destroy the periodic atomic structure of the crystal (so acceleration will take place only in a short time before full dissociation of the lattice). For the fields of about $\SI{1}{\giga\volt\per\centi\meter} = \SI{0.1}{\tera\volt\per\meter}$ or less, reusable crystal accelerators can probably be built which can survive multiple pulses. 

The introduction of crystals and nanotubes into compact accelerators has been motivated by several directions.  One was induced by the potential emergence of a new X-ray laser technology \cite{MOUROU2014} that allows the optical laser conversion into an X-ray laser, thus the operating regime of laser-driven wakefields can enter much higher density than gas \cite{TAJIMA2014}. It is also driven by the progress in obtaining far shorter and denser electron bunches than with the previously available techniques, exemplified by the FACET-II facility~\cite{FACETII} at SLAC. Furthermore, schemes were proposed, e.\,g., for bunch compression down to an attosecond-scale~\cite{Emma2021}. Such ultra-short electron bunches can now interact with much higher density medium such as nanotubes \cite{IIJIMA} (or even solids). The flexibility to choose the average electron density of nanotubes also introduces a possibility to adjust for resonant wakefield excitation at various pulse lengths, dielectric variabilities, and geometries that may be controlled by the emerging nanotechnology.  For example, hollow nanomaterials with parameters suitable for acceleration may be manufacturable soon. 
 
\begin{figure}[ht]
\begin{center}
\includegraphics[width=5.6in]{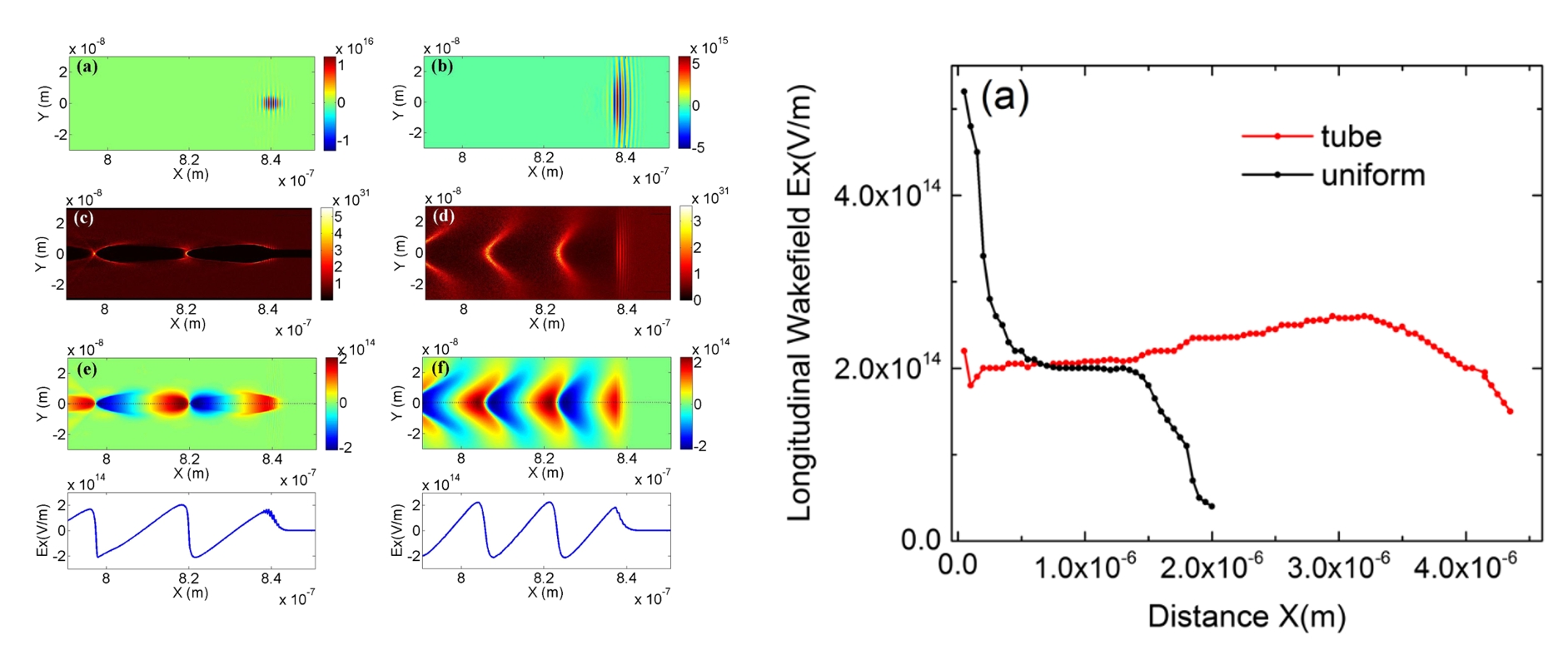}
\end{center}
\caption{Left - Wakefield excitation in a tube by X-ray laser pulse (left colum of plots) in comparison with a wakefield in a uniform system (right column). Shown are
distributions of the laser field (a, b), electron density (c, d), and  longitudinal wakefield (e, f). X-ray pulse spot size of \SI{5}{\nano\meter} and length of \SI{3}{\nano\meter}. Right - in the case of amorphous media the laser field (and wakefields) decreases rapidly with the propagation distance. In the nanotube, the X-ray pulse maintains a small spot size and propagates over much longer distance generating strong nonlinear wakefields (from Ref. \cite{ZHANG2016}).}
\label{zhangfig2}
\end{figure}
 
Ref.~\cite{ZHANG2016} examined the effects of hollow nanotube structures vs.\ uniform materials for X-ray-driven laser wakefield acceleration. Figure~\ref{zhangfig2} compares the wakefield intensity, its shape and stability, and the electron dynamics by these wakefields.  With the hollow nanostructure, the electron acceleration and dynamics are found to be superior.  Also, it is evident that the accelerating gradients are ultra-high, as expected due to correspondingly higher electron density of the materials that supports the wakefield -- per Eq.~(\ref{grad}).  
Previous simulation studies of wakefields driven by short electron bunches \cite{SAHAI2020} indicate similar behavior as that reported in~\cite{ZHANG2016}. In addition, one can be interested in the properties of nanotubes (and nanomaterials) due to their ionic structures studied in~\cite{HAKIMI2018}. The wakefield properties are found to be generally similar even if the optical phonon effects due to the lattice structure are incorporated~\cite{TAJIMA1978}.

The concept of crystal colliders is yet very hypothetical and various requirements like the generation of suitable drivers, power efficiency, and the asymmetric acceleration of muons and antimuons need to be addressed. In the following two sections, we will discuss near-term experiments that may already allow to address some of these issues with currently available technology and facilities and the prospects of mid-term facility upgrades. The fourth section gives an outlook of channeling acceleration and its use in various collider applications.

\section{Near-term prospect for beam solid-plasma experiments}
\label{near-term-exp}

\begin{figure}[ht]
    \centering
    \includegraphics[width=0.95\textwidth]{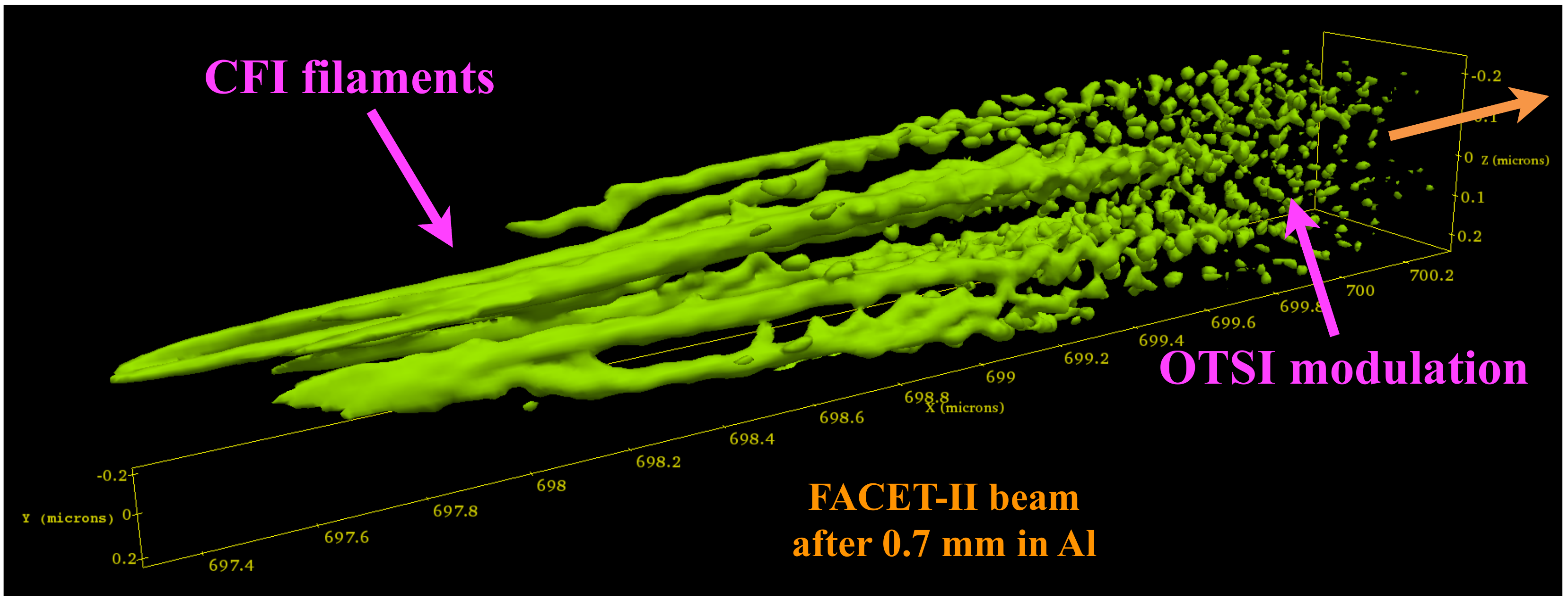}
    \caption{3D PIC simulation of the \SI{10}{\giga\electronvolt} FACET-II electron beam interacting with an Al solid target. After \SI{0.7}{\milli\meter} of propagation, strong oblique modulations (OTSI) have built up at the beam front (right), which evolve into transverse filaments (CFI) when moving toward the rear of the beam (left). The simulation was run using the \textsc{calder} 3D PIC code~\cite{CALDER}, with a peak beam density $n_b=\SI{1e20}{cm^{-3}}$, a normalized transverse emittance of \SI{3}{\milli\meter\milli\radian}, a bunch length of \SI{1}{\micro\meter} and with periodic boundary conditions in the transverse directions.}
    \label{fig-e305solid}
\end{figure}

Although the ultimate conditions to drive wakefields in nanostructures or crystals require transverse and longitudinal sizes comparable or smaller than the plasma wavelength $\lambda_p$, which ranges from \SIrange{30}{300}{\nano\meter} for densities $n_p$ from \SIrange{e22}{e24}{\per\cubic\centi\meter}, first experimental tests can be carried out before these conditions are reached. This is the purpose of the E-336 experiment that is planned to be conducted with the FACET-II accelerator facility~\cite{FACETII} at SLAC National Accelerator Laboratory, where the driver is an electron beam with transverse and longitudinal sizes expected to be in the range of \SIrange{1}{10}{\micro\meter}. Together with the \SI{2}{\nano\coulomb} beam charge, \SI{10}{\giga\electronvolt} beam energy and low \SI{\sim 10}{\milli\meter\milli\radian} normalized emittance, such beams are suitable to study the physics developing at the scale of $\lambda_p$ when the electron beams propagate and interact with solid media that are either amorphous (E-305 experiment) or nanostructured (E-336 experiment).

In an amorphous solid both the longitudinal and transverse size of the beam are large compared to $\lambda_p$. Thus, wakefield excitation is very inefficient and wakefields can be damped by collisional effects. What is left to the plasma response is the return current that flows through the beam, and leads to an unstable counterstreaming system of beam and plasma electrons, where any perturbation at the scale of $\lambda_p$ can grow exponentially as the beam propagates through the solid. For ultrarelativistic beams such as delivered by FACET-II, two modes of instability are prominent: the oblique two-stream instability (OTSI)~\cite{SANMIGUEL2021}, which is mainly electrostatic and with a wave vector at an oblique angle with respect to the beam propagation, and the current filamentation instability (CFI), which is mainly magnetic and transverse. This is exemplified in Fig.~\ref{fig-e305solid} showing the FACET-II electron beam after \SI{0.7}{\milli\meter} propagation through aluminum, as simulated using the \textsc{calder} 3D PIC code~\cite{CALDER}. At the front of the beam (right of Fig.~\ref{fig-e305solid}), longitudinal and transverse modulations are clearly observed and correspond to the OTSI instability, which is the fastest mode to grow in the linear phase~\cite{SANMIGUEL2021}. In contrast, at the rear of the beam, beam filaments can be identified and attributed to the CFI instability. The interplay between these two modes that appear respectively at the beam front and rear of the beam can be understood by the variation of the collisionality as the plasma electrons are heated when flowing from the front to the rear of the beam. In this simulation, the higher collisionality at the rear damps the OTSI and favors the CFI instability, thus resulting in the beam profile shown in Fig.~\ref{fig-e305solid}). Beyond the imprint that the beam-solid interaction has left on the beam at a scale of $\lambda_p$, high-energy beam electrons also experience the very large electromagnetic fields associated with the instability and thus can radiate gamma rays very efficiently~\cite{BENEDETTI2018}, and such gamma-ray flashes can also be an excellent experimental observable to diagnose the beam-solid interaction and the physics developing at the scale of $\lambda_p$.

\begin{figure}[ht]
    \centering
    \includegraphics[width=0.95\textwidth]{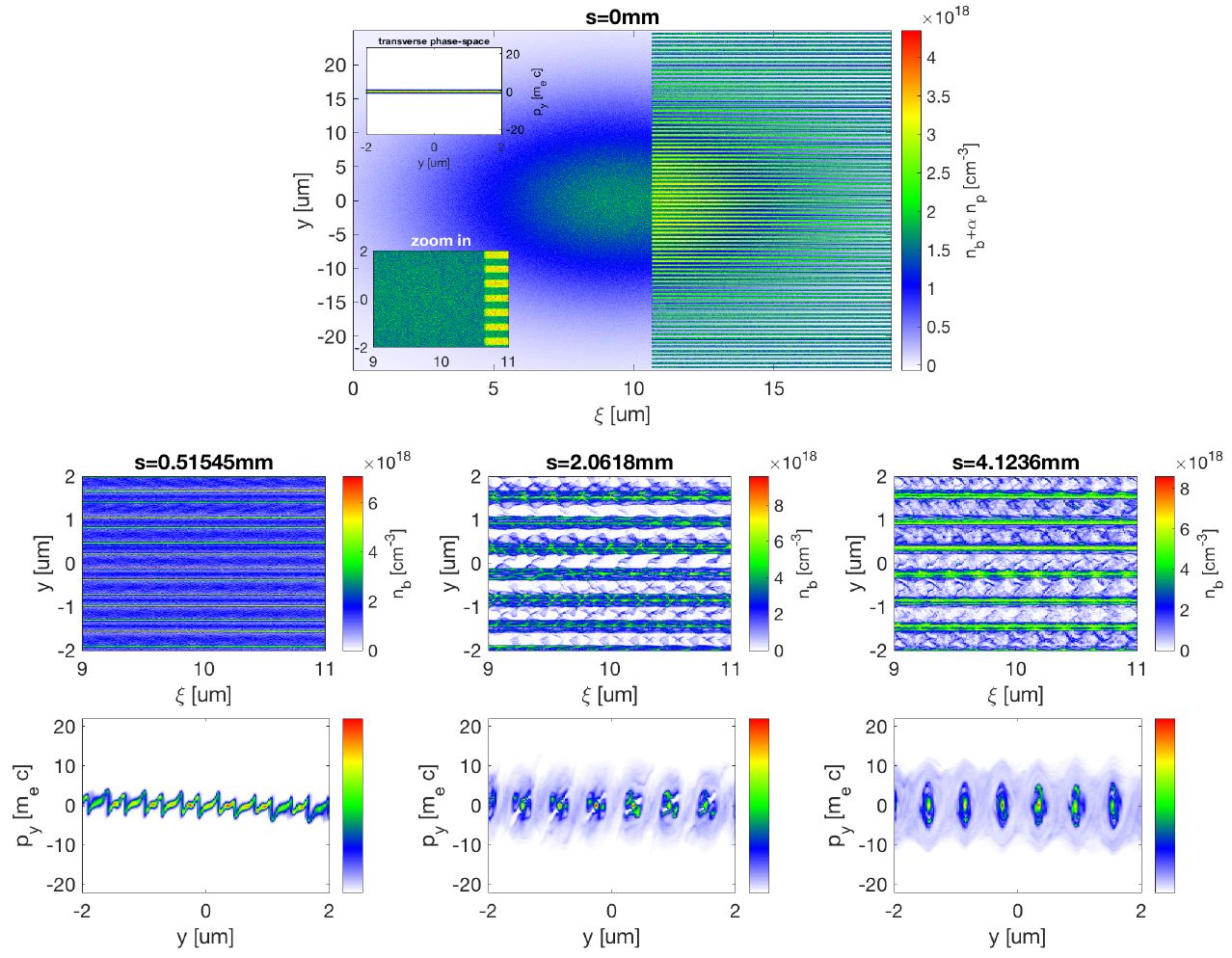}
    \caption{2D PIC simulation of the \SI{10}{\giga\electronvolt} FACET-II electron beam interacting with a nanostructured target. The target (see top figure) consists in 300-\si{\nano\meter}-wide vacuum gaps separated by 300-\si{\nano\meter}-wide plasma sections of electron density $n_p=\SI{2e22}{\per\cubic\centi\meter}$. The beam density (middle row) and transverse phase space (bottom row) are shown after different propagation distances ($s=\SI{0.5}{\milli\meter}, \SI{2.1}{\milli\meter}, \SI{4.1}{\milli\meter}$). The simulation was run using the \textsc{calder} 2D PIC code~\cite{CALDER}, with a beam charge of \SI{2}{\nano\coulomb}, a peak current of \SI{50}{\kilo\ampere}, a normalized transverse emittance of \SI{5}{\milli\meter\milli\radian} and a beam size of \SI{10}{\micro\meter}.}
    \label{fig-e336}
\end{figure}

In a nanostructured solid, the beam evolution is no longer dictated by the amplification of initial noise perturbations that can exist at the scale of $\lambda_p$, but by the imposed structure of the solid. Notably, for targets with hollow tubes where the beam is much larger than $\lambda_p$ and the structure length scale (e.\,g.\ the tube diameter), it fills both the vacuum and solid parts of the structure, and many periods transversely (see Fig.~\ref{fig-e336}). While the beam self fields can exist in the vacuum tubes of the nanostructure, they are shielded in the solid which results in the excitation of wakefields and to the channeling of beam electrons in the vacuum tubes. The physics also differs depending on the typical length scale of the structure. For a tube diameter much larger than $\lambda_p$, shielding is done over a skin depth that is much smaller than the tube diameter and thus the solid plasma response is mainly driven on the tube surface. In contrast to this surface response, for tube diameter of the order of $\lambda_p$ or smaller, the response of plasma electrons can become more volumetric.

\begin{figure}[ht]
\begin{center}
\includegraphics[width=\textwidth]{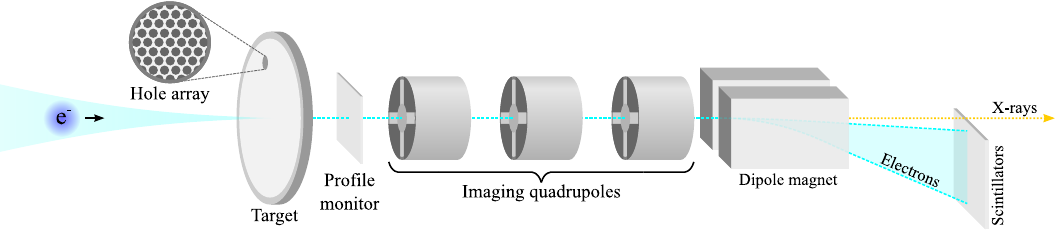}
\end{center}
\caption{Sketch of the E-336 experimental setup at FACET-II. The electron bunch is focused on a target with a hole array. The beam after the interaction passes through an optional profile monitor which allows to measure the transverse momentum spread and deflection. It is thereafter imaged with a quadrupole triplet onto a scintillating screen. A dispersive dipole magnet allows the reconstruction of the energy spectrum. The X-rays produced in the beam-target interaction are measured with profile monitors and spectrometers (not shown).}
\label{experimental_setup}
\end{figure}

A typical outcome of beam-nanotarget interaction is depicted in Fig.~\ref{fig-e336} for a beam whose size and length is very large compared to the nanostructure period and to $\lambda_p$. The density of the \SI{10}{\giga\electronvolt} beam after a \si{\milli\meter}-scale propagation (see middle row of Fig.~\ref{fig-e336}) clearly exhibits a transverse nanomodulation at the nanostructure period with density maximums located in the vacuum gaps. The transverse phase space of the beam (see bottom of Fig.~\ref{fig-e336}) shows that the transverse momentum spread, and thus the angular spread, is greatly increased due to the beam-nanotarget interaction, and it takes the typical shape (see bottom right of Fig.~\ref{fig-e336}) associated to trapped beam electron orbits (around the peaks visible in the phase space). This demonstrates the channeling principle with beam electrons being captured and trapped in the vacuum gaps. Beyond nanomodulations and growth of angular spread, an additional experimental observable is the generation of radiation, similarly to the case of amorphous solid. 
Indeed, the channeling of beam electrons in the vacuum tubes leads to the emission of betatron-like radiation by beam electrons oscillating about the axis of the tube. The radiation frequency can provide a direct measurement of the particle oscillation period and thus of the focusing force in the tube. Figure~\ref{experimental_setup} shows a sketch of the experimental setup. The beam energy distribution after the interaction is measured with a dipole-based electron spectrometer combined with an imaging quadrupole triplet, and the generated radiation is measured with X-ray profile monitors and spectrometers.

\begin{figure}[ht]
    \centering
    \includegraphics[width=0.95\textwidth]{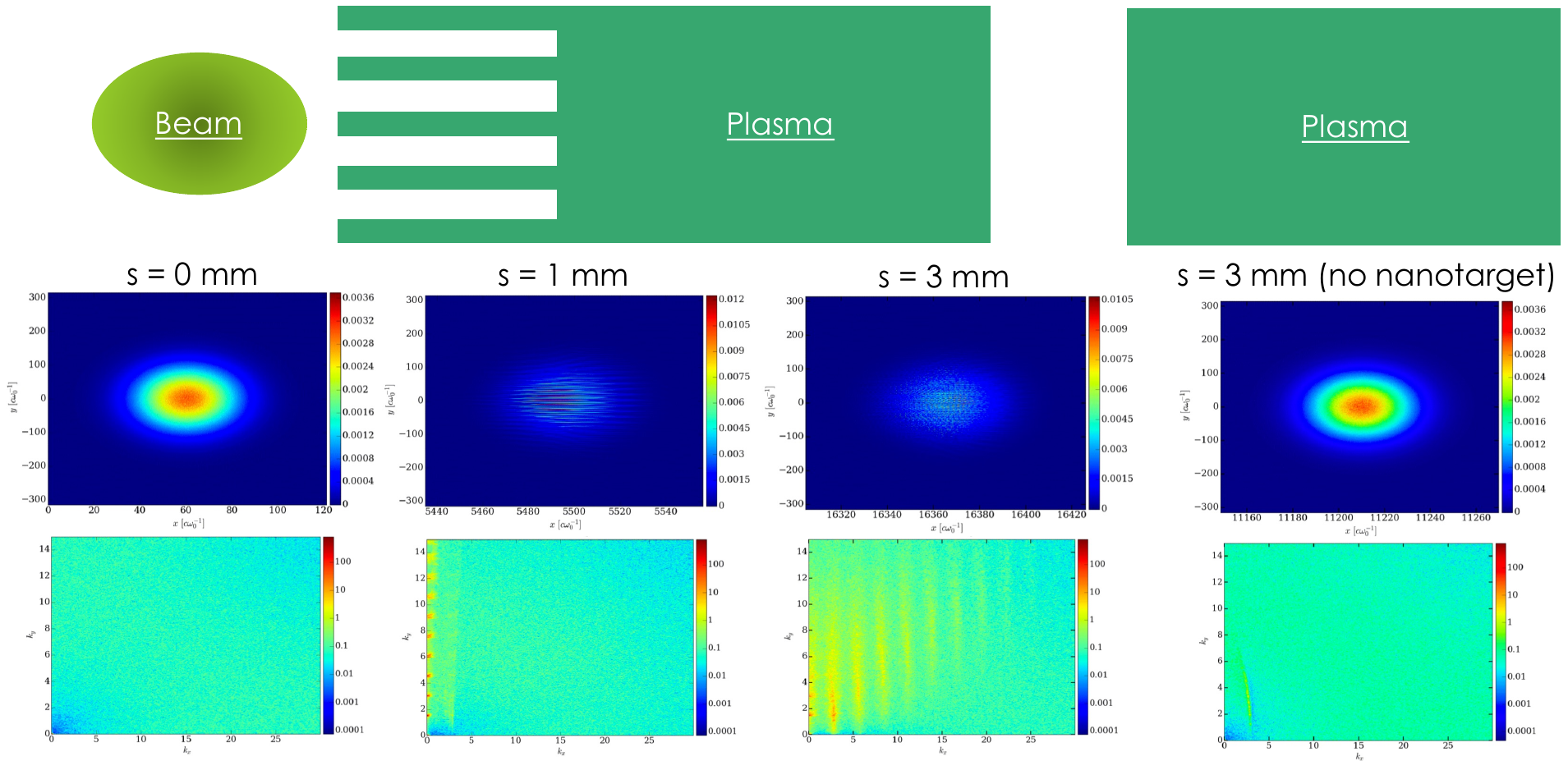}
    \caption{2D PIC simulation of the \SI{10}{\giga\electronvolt} FACET-II electron beam interacting with a 1-mm-long nanostructured target followed by a 2-\si{\milli\meter}-long uniform plasma (see top left). The beam density (middle row) and its Fourier transform (bottom row) are shown after different propagation distances corresponding to the start of the simulation ($s=\SI{0}{\milli\meter}$), the end of the nanotarget ($s=\SI{1}{\milli\meter}$) and the end of the uniform plasma ($s=\SI{3}{\milli\meter}$), as well as for the case without nanotarget (right column, 2-\si{\milli\meter}-long uniform plasma only).}
    \label{fig-e336_seed4e305}
\end{figure}

Another new opportunity opened by the study of the response of solid plasmas to extreme beams is to take advantage of both amorphous and nanostructured solids. An interesting geometry is to start with a nanostructured target, that can be understood as a pre-modulator for the beam, and then go into an amorphous solid that can work as an amplifier where instabilities will take place and leave a strong imprint on the beam and on the level of gamma-ray emission. Such scenario is of interest first because the nanotarget induces a nanomodulation that acts as a seed for the instability in the amorphous solid, but also because the experimental observation of the instability after the nanotarget can reveal very effectively that a process of nanomodulation indeed took place in the nanotarget. 
This strategy thus allows us to be extremely sensitive to the beam evolution in the nanotarget. The principle is illustrated in Fig.~\ref{fig-e336_seed4e305} by a \textsc{calder} 2D PIC simulation. In conditions for which no instability is observed in a purely uniform plasma (see right column of Fig.~\ref{fig-e336_seed4e305}), the addition of a 1-\si{\milli\meter}-long nanostructured target induces a transverse modulation corresponding to the harmonics of the spatial frequency of the nanostructure (see second column corresponding to $s=\SI{1}{\milli\meter}$ in Fig.~\ref{fig-e336_seed4e305}) and appearing only at $k_\|=0$. After going through the following 2-\si{\milli\meter}-long uniform plasma, the pre-modulation from the nanotarget allows the beam to undergo the instability with a very strong longitudinal modulation characteristic of the OTSI (with peaks in the Fourier transform when $k_\|$ is a multiple of $k_p$) and a transverse modulation that spans $k_\perp$ continuously in the Fourier transform (instead of being multiple of the nanostructure spatial frequency). Because this strong instability-driven beam modulation (see bottom row and third column of Fig.~\ref{fig-e336_seed4e305}) is only present when there is a nanostructured target to pre-modulate the beam, this simulation result illustrates how this scheme can provide an experimental signature that is very sensitive to the nanomodulation from the nanostructured target. For instance, transition radiation emitted as the beam exits the uniform plasma will exhibit the longitudinal structure of the beam, and its spectrum will thus reveal the longitudinal beam modulation that is reminiscent of the nanomodulation from the nanostructured target.

\section{Paths and facilities for follow-up experiments}
\label{future-exp}

To go beyond these near-term experiments, novel developments and facilities are required. Indeed, ultimately we need drivers (X-ray pulses, electrons or high-Z ions) with transverse and longitudinal sizes comparable or shorter than the plasma wavelength $\lambda_p$ at solid density, that is of the order of \SI{0.1}{\micro\meter}. A first objective should be to develop a new capability in existing accelerator facilities such as FACET-II, that is to deliver electron beams focused transversely to sub-$\mu$m beam size.
Thin plasma lenses are excellent candidates to readily focus a beam that is initially few \si{\micro\meter} in size down to sub-\si{\micro\meter} sizes~\cite{Doss2019}, and they open a promising path to reach beam sizes of the order of \SI{0.1}{\micro\meter} that are relevant for crystal/nanostruture wakefield acceleration. The successful implementation of such thin plasma lenses in an accelerator facility will thus be a great asset to study wakefield excitation in crystals and nanostructures. Such highly-focused beams could also come from novel high-brightness beam sources such as the plasma photocathode~\cite{HIDDING2012}, or laser-based schemes by leveraging the capabilities of ultrahigh-power laser systems. The latter can not only provide new sources of electron beams, but also of coherent X-ray pulses.

With this prospective availability of \textbf{highly-focused} but still \textit{long} (compared to $\lambda_p$) drivers, one can already envision to study crystal/nanostructure wakefield. This can first be achieved by relying on self modulation due to the wakefields in the nanostructure, in a way similar to proton bunch self modulation observed in the AWAKE experiment~\cite{ADLI2018, TURNER2019}. One can also decouple the process of beam modulation and the excitation of wakefield in the nanostructure by sending a pre-modulated driver into the structure. This could be achieved for instance through self modulation of the beam in an amorphous solid that is preceding the nanostructure. Such advanced experiments will beneficiate from the input of near-term experiments such as E-336 at FACET-II, that will provide valuable experimental measurements allowing us to benchmark our theoretical modeling of beam-nanostructure interaction against experimental data. Advanced nanostructure experiments with highly-focused drivers are expected to provide a unique insight into the interaction of ultrahigh-density drivers (in particular electron beams) with crystals and different types of nanostructures such as CNTs and porous alumina or glass nanostructures. It will further give valuable experimental data on controlled focusing and self-bunching/slicing/modulation of ultrahigh-density beams in CNTs/crystals, and on the generation and detection of high-amplitude wakefields through the coherent buildup from a pre-modulated beam.

The most challenging development is arguably to deliver {\bf highly-compressed drivers}, first at sub-\si{\micro\meter} bunch length and ultimately reaching the order of \SI{0.1}{\micro\meter}. While accelerator facilities such as FACET-II may be able to deliver $\sim\si{\micro\meter}$ bunch lengths with peak currents from \SIrange{100}{300}{\kilo\ampere}~\cite{FACETII}, the experimental demonstration of a \SI{0.1}{\micro\meter} bunch length requires novel developments. This could be done directly through some specific plasma-based and/or laser-based injection schemes (e.\,g.~\cite{HIDDING2012}), or by re-compressing beams produced in plasma and/or laser wakefield accelerators~\cite{Emma2021}. 
Although the characterization of the longitudinal phase space of electron beams from plasma accelerators is still in its infancy, theoretical and numerical modeling suggests that the longitudinal beam quality (in particular the longitudinal emittance) is excellent and the energy spread is actually dominated by a correlated energy spread. Manipulating the longitudinal phase space and re-compressing electron beams from plasma accelerators is thus a very promising approach to reach \SI{0.1}{\micro\meter} bunch length with mega-ampere peak currents~\cite{Emma2021}. For X-ray drivers, the recent invention of the thin film compression technique promises to deliver relativistically compressed X-ray laser pulses, which are suited for crystal/nanostructure channeling acceleration~\cite{Mourou2014,Naumova2004}. The combined availability of \textbf{highly-focused and highly-compressed drivers} will mark a critical milestone, allowing to considerably push the research on crystal/nanostructure wakefield acceleration with its full capability unravelled, that is with the generation of extreme-gradient (\si{\tera\electronvolt\per\meter}) wakefields and possibly to coherent sources of X-rays and gamma rays. We should emphasize that such an incredible development with extreme highly-focused and highly-compressed drivers will benefiate to the physics community well beyond crystal/nanostructure wakefield acceleration, for example it will be a critical stepping stone on the way to a fully non-perturbative QED collider~\cite{YAKIMENKO2019} and will be instrumental to novel schemes that could allow for extremely dense gamma-ray and $e^+e^-$ pair jets and laserless study of strong-field QED~\cite{SAMPATH2021}, both of great interest for our fundamental understanding of the laws of nature and for astrophysics.

\section{Crystal channeling principles and collider applications}
\label{collider}

Channeling in crystals has many promising applications in the development of future accelerators and colliders. Beyond the acceleration itself, these applications include beam steering and focusing, as well as the generation of charged particle beams. 

\begin{figure}[t]
\begin{center}
\includegraphics[width=3.8in]{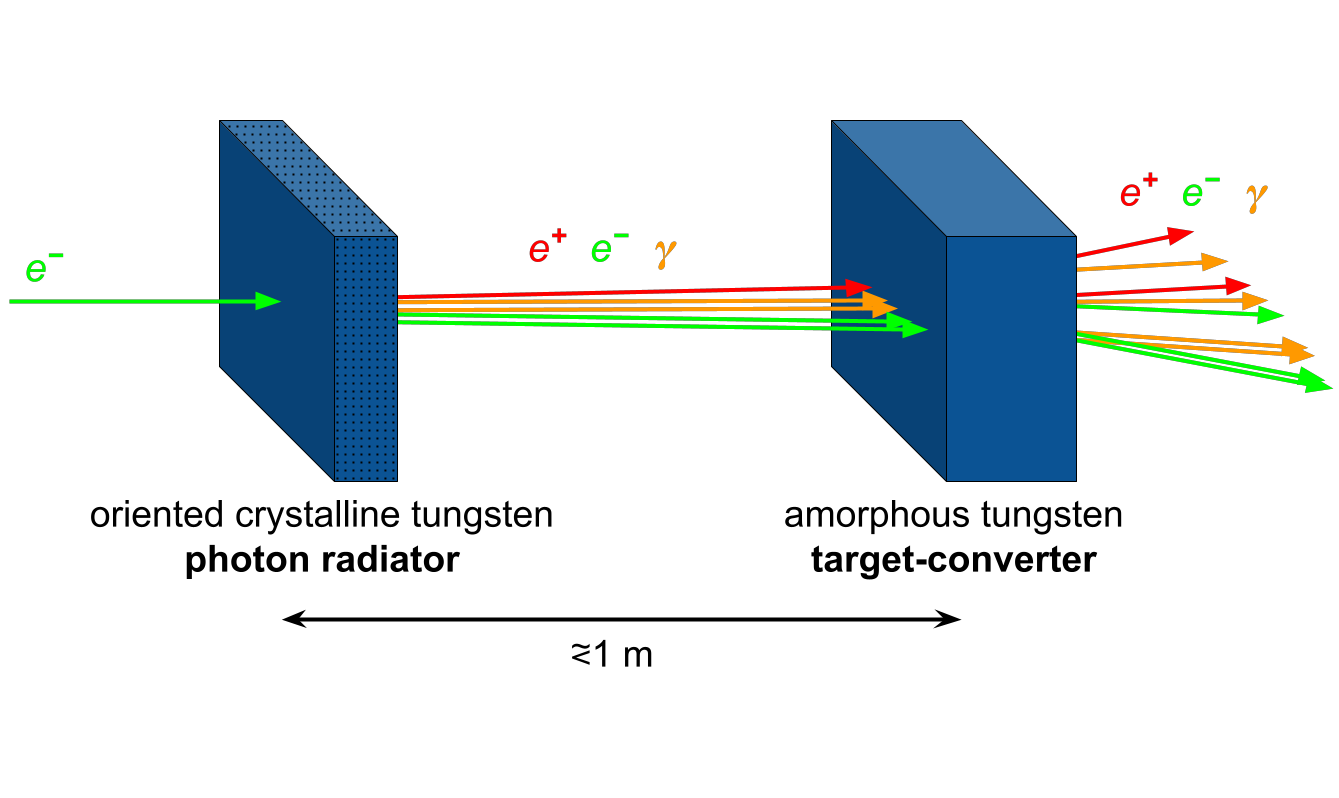}
\end{center}
\caption{Schematic representation of a hybrid crystal-based positron source~\cite{Bandiera2022}.}
\label{positron_source}
\end{figure}

In particular, a heavy crystalline material like tungsten is very promising for a hybrid crystal-based positron source~\cite{Bandiera2022,Soldani2023} of future $e^+ e^-$ colliders such as FCC-ee or ILC. The basic idea is to exploit two targets, a $\gamma$-quanta radiator (oriented W crystal) and a subsequent gamma-to-positron converter as shown in Fig.~\ref{positron_source}. This allows amplifying the radiation intensity in the crystalline radiator using channeling radiation and coherent bremsstrahlung effects, and therefore increasing the positron yield. At the same time the enlarged beam spot on the second target reduces its radiation load, being the main limitation of the positron source. Moreover, if these positrons are accelerated to above \SI{\sim 44}{\giga\electronvolt} (which is the $e^+ e^- \Rightarrow \mu^+ \mu^-$ conversion threshold) they can produce very low emittance muon beams for a $\mu^+ \mu^-$-collider as proposed in the LEMMA project~\cite{Alesini2019}.

\begin{figure}[t]
\begin{center}
\includegraphics[width=3in]{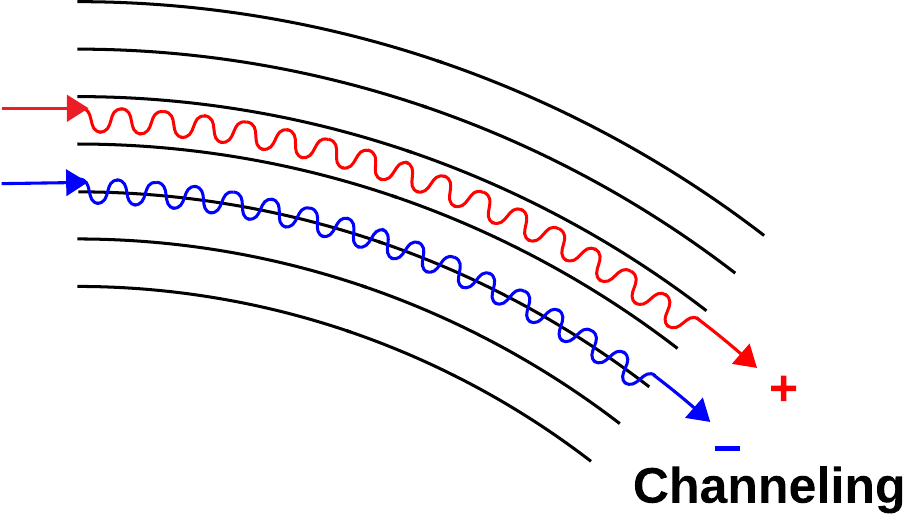}
\end{center}
\caption{Schematic representation of channeling in a bent crystal. The red (+) and blue (-) curves indicate channeling trajectories of positively and negatively charged particles, respectively. Black lines indicate crystal planes.}
\label{bent_crystal}
\end{figure}

Channeling in bent crystals proposed by Tsyganov~\cite{Tsyganov1976} has a promising steering capability (as shown in Fig.~\ref{bent_crystal}), e.\,g., for directing beams into an IP or target. The strong electric fields of the crystal planes and atomic strings allow deflecting charged particles equivalently to magnetic fields of more than \SI{100}{\tesla}.
Crystals with a varying bending angle allow focusing charged particle beams (as experimentally demonstrated in~\cite{Denisov1992,Biryukov2005,Scandale2014}), which makes them a promising candidate for the beam focusing in a collider IP.
Furthermore, bent crystals have already been successfully used for collimation and extraction at different hadron machines including U-70, SPS, Tevatron and LHC~\cite{Afonin2012,Scandale2016,Mokhov2010,Scandale2012}. They are currently considered as a baseline solution for the ion beam collimation of the HI-LUMI LHC~\cite{Apollinari2014} and are considered as potential collimators for future muon colliders. 

The channeling process is particularly efficient for positively charged particles, as they are repelled from the crystal nuclei and thus propagate in between the crystal planes. On the other hand, negatively charged particles cross the crystal planes at every channeling oscillation which considerably reduces their channeling efficiency due to multiple scattering. However, recent experiments at CERN, MAMI and SLAC~\cite{Sytov2017,Mazzolari2014,Bandiera2015,Bagli2017,Wienands2015,Wistisen2017} still show rather high channeling efficiencies of \SIrange{20}{40}{\percent} for electrons. 
Channeling of muons is potentially more efficient since they do not experience nuclear interactions as hadrons and have less radiation losses than electrons/positrons.

\begin{figure}[t]
\begin{center}
\includegraphics[width=5in]{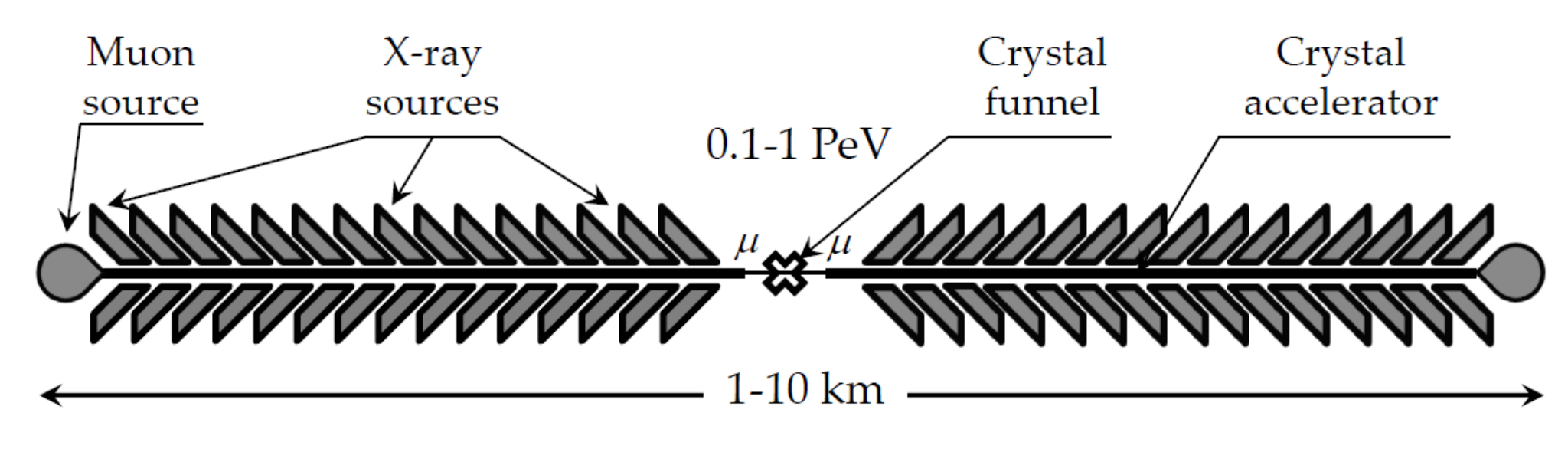}
\end{center}
\caption{Concept of a linear X-ray crystal muon collider (adapted from \cite{VS2012}).}
\label{crystal_collider}
\end{figure}

Combining these techniques with the crystal acceleration capabilities could serve as novel linear muon collider. A conceptual scheme of such a collider is shown in Fig.~\ref{crystal_collider}. It includes two high brightness muon sources, two continuous crystal linacs of a total length of \SIrange{1}{10}{\kilo\meter} driven by numerous X-ray sources (or other type of drivers) and could reach \SIrange{1}{10}{\peta\electronvolt} c.m.e.\ at the interaction point~\cite{VS2012}. 
Initial luminosity analysis of such machine assumes the minimal overlap area of the colliding beams to the crystal lattice cell size 
$A \sim \SI{1}{\angstrom} = \SI{e-16}{\per\square\centi\meter}$ and that the crystals in each collider arm are aligned channel to channel. The number of muons per bunch $N$ also can not be made arbitrary high due to the beam loading effect and should be $N \sim 10^3$. Exciting many parallel atomic channels $n_{\mathrm{ch}}$ can increase the luminosity $L=fn_{\mathrm{ch}} N^2 A^{-1}=10^{16} \cdot 10^6 \cdot n_{\mathrm{ch}}[\si{\per\square\centi\meter\per\second}]$ which can reach \SI{e30}{\per\square\centi\meter\per\second} at, e.\,g., $f = \SI{e6}{\hertz}$ and $n_\mathrm{ch} \sim 100$. Exceeding the value of the product $fn_\mathrm{ch}$ beyond 
\SI{e8}{\hertz} can be very costly as the total beam power $P = f n_\mathrm{ch} N E_p$ will get beyond \SI{\sim 10}{\mega\watt}. Instead, using some crystal focusing to bring microbeams from many channels into one can increase the luminosity by a factor of $n_\mathrm{ch}$ to some \SI{e32}{\per\square\centi\meter\per\second}.

\section{Conclusions}

The acceleration in crystals and nano-structured targets like CNTs offers order-of-magnitude higher acceleration gradients than the current state-of-the-art RF technology and even gas-based plasma wakefield accelerators. However, effective excitation of the wakefields in the solid density plasmas requires short and ultra-intense drivers not commonly available at present. Unmatched high energy beam properties of the FACET-II accelerator facility at SLAC~\cite{FACETII} offer unique opportunities for the initial proof-of-principle experimental studies of the wakefields and acceleration in structured solid targets.

The E-336 experiment at the FACET-II aims at studying the interaction of micrometer-size electron beams with amorphous and nanostructured targets, which may give a better understanding of the physics down to a level of solid-density $\lambda_p$, which is typically in the range of \SIrange{30}{300}{\nano\meter}.
In near-term experiments, solid glass targets with nanometer to micrometer-size holes are used where the impinging electron beam spans over multiple holes at once. Simulations show a transition from strong interaction on the hole surfaces for \si{\micro\meter}-size holes to volumetric effects when the hole size approaches the plasma skin depth. In both of these regimes, the beam experiences a transverse modulation that qualitatively and quantitatively differs with respect to the hole size. Foremost, this modulation essentially converts the relatively large initial beam into beamlets with sizes possibly down to the nanometer scale. These measurements will thus yield valuable physics of the beam-nanostructure interaction on different scales and possibly down to solid $\lambda_p$ beam size.

The strong transverse modulation imposed by the structured targets has promising applications, e.\,g., the study of filamentation instabilities in beam-solid interactions. Here, the modulation offers an unprecedented degree of control as it can be used to seed the filamentation with desired spatial frequencies and strengths.

Beyond the near-term experiments, either novel accelerator facilities or mechanisms like thin plasma lenses or plasma photocathodes may open the path towards strongly focused drivers of transverse solid density $\lambda_p$ sizes ($\sim$ \SIrange{30}{300}{\nano\meter}). These can be used to drive crystal/nanostructure accelerators in a self modulated regime. Techniques like the modulation and subsequent recompression of electron beams or thin film compression for X-ray laser pulses may furthermore also bring the driver length towards $\lambda_p$, such that the drivers can efficiently be used to drive the acceleration process. Such highly-focused and highly-compressed beams will be a substantial milestone in the physics community for various fields.

Crystals and nanostructured targets do not only offer ultra-high acceleration gradients, but also the use as various other beam line elements. Crystal planes can exert fields equivalent of more than \SI{100}{\tesla} on beam particles passing through. Bent crystals can be used for steering and collimating beams, and for a tight focusing into the IP of a collider. Ultimately, crystals and nanostructured targets with their acceleration and steering/focusing capabilities may offer a promising path towards highly anticipated future linear muon colliders.

\acknowledgments
The work at LOA and CEA was supported by the ANR (UnRIP project, Grant No. ANR-20-CE30-0030). The work at LOA was also supported by the European Research Council (ERC) under the European Union's Horizon 2020 research and innovation programme (M-PAC project, Grant Agreement No. 715807). A. Sytov acknowledges support by the European Commission through the H2020-MSCA-IF TRILLION project (GA. 101032975). H. Piekarz and V. Shiltsev's work was supported by the Fermi National Accelerator Laboratory, managed and operated by Fermi Research Alliance, LLC under Contract No. DE-AC02-07CH11359 with the U.S. Department of Energy. The work at SLAC was supported by U.S. DOE FES Grant No. FWP100331 and DOE Contract DE-AC02-76SF00515.

\end{document}